# Hard X-Ray Phase-Contrast Laboratory Microscope Based on Three-Block Fresnel Zone Plate Interferometer

**L.A. Haroutunyan**

*Yerevan State University, Yerevan, Armenia*
*levon.har@gmail.com*

A device based on a three-block Fresnel zone plate interferometer is proposed for hard X-ray phase-contrast imaging. The device combines a low requirement for the coherence of the initial radiation (the interferometer operates in the amplitude division mode) with an optical magnification of the image. A numerical simulation of the image formation is carried out, taking into account the limited source–interferometer distance, the size and spectral width of the X-ray source. The calculations show that the proposed set-up can be used as a phase-contrast microscope using laboratory hard X-ray sources.

## 1. Introduction

The first implementation of the interferometric method of hard X-ray phase-contrast imaging is based on the so-called triple Laue-case (LLL) interferometer [1, 2]. The tested phase object (an object made of light elements), is located in the second interblock space, across one of the interfering beams of the object wave. This leads to the phase modulation of the object wave, which affects the location and shape of the interference fringes arising in the interferometer [3]. The reconstruction of the phase-shift spatial distribution of the object wave from the registered interference pattern can be carried out by one of the following methods: the "Fourier-transform method" [4] or the "fringe scanning method" [5]. However, for objects with small enough phase shifting, direct mapping can be used [6]. The interferometer operates in amplitude-division mode and doesn't impose strong requirements on the coherency of the initial radiation. This feature made its experimental realization possible in the 60s of the last century using the hard X-ray laboratory sources of those years [1].

Fresnel zone plates (FZP) are widely used in X-ray optics, both for radiation focusing and for imaging [7]. As diffraction gratings, they possess different diffraction orders. This significantly reduces the focusing efficiency and causes problems in image formation. But, on the other hand, it allows the use of FZP in interferometry as wave splitters and analyzers. This feature allows the combination of image magnification with the wave interference still in the step of image formation.

Experimental set-ups for X-ray phase contrast imaging in the optical magnification mode, based on interferometers with two FZPs, have already been proposed [8–10]. However, these interferometers are wavefront-division type, and, consequently, impose strong requirements for the coherency of initial radiation. Nevertheless, these requirements are satisfied by the contemporary third-generation synchrotron radiation facilities.

A three-block FZP interferometer, operating in amplitude division-mode and not imposing strict requirements on the coherency of initial radiation, has already been presented by the author of this work [11]. The possibility of application of this interferometer for phase-contrast imaging in the case of the initial plain-monochromatic radiation has been shown by a numerical simulation [6].

A similar device, but for hard X-ray laboratory sources is considered in this paper. For this purpose, the modification of the mentioned interferometer for a limited distance radiation source–interferometer is proposed. The possibility of phase-contrast imaging with the modified interferometer, using the monochromatic point-source initial radiation is considered by a numerical simulation. The case of initial radiation with limited temporal and spatial coherency, which better approaches the laboratory sources, is also considered.

## 2. Three-block FZP interferometer for initial plain-wave radiation

The interferometer under consideration consists of three similar FZPs. The FZPs have a common optical axis, the distances between adjacent blocks are equal to the doubled focal length of the first-order diffraction of the FZPs (see. Fig. 1). The initial plane wave falls parallel to the optical axis on the first FZP and the interference pattern is registered behind the third FZP, in a plane perpendicular to the optical axis. Taking into account only 0, +1, and –1 orders of diffraction on the FZPs, 27 propagation channels (PC) are formed in the interferometer. The interference between the following two PCs is considered: (a) the beam diffracted in order +1 on the first and second and in order 0 on the third FZP, and (b) the beam diffracted in order 0 on the first and in order +1 on the second and third FZPs. The suppression of the influence of the remaining PCs on the registered interference pattern is carried out by two knife-edges located on the first and third FZP, as shown in Fig. 1. In the geometrical optics approach, it can be shown by tracing the rays of PCs, that the mentioned "unwanted" PC does not intersect with the registered interference pattern if the detector distance from the third FZP ($f$) is large enough:

$$f > \max(5/3\,, R/d - 1)F. \quad (1)$$

Here $F$ and $R$ are: the focal length of the first diffraction order and the radius of the FZPs, respectively, and $d$ is the distance of the knife-edges from the optical axes of the interferometer.

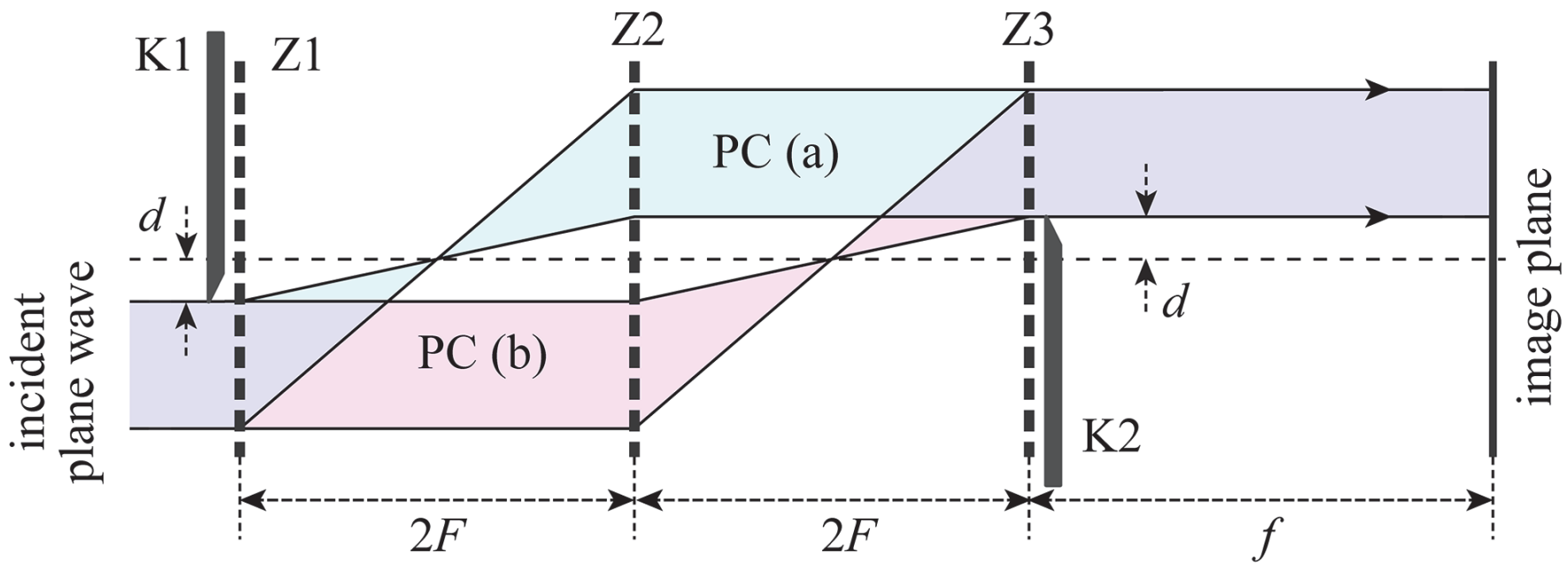

**Fig. 1.** Schematic diagram of the three-block interferometer for an initial plane wave. Z1, Z2, and Z3 are the interferometer blocks, K1 and K2 are the knife edges.

Note that these knife-edges, in addition to suppressing "unwanted" PCs, also provide spatial separation of PCs (a) and (b), which is important in the case of phase-contrast imaging.

As can be seen from Fig. 1, the first block of the interferometer (first FZP) acts as a splitter, splitting the initial plane wave into converging (PC (a)) and parallel (PC (b)) beams. The diffraction of +1 order was used in the first case, and 0 order – in the second case. The second block (mirror) converts the diverging beam of PC (a) into a parallel one, and the parallel beam of PC(b) into a converging one (in both cases the +1 diffraction order takes place). Finally, the third block (analyzer) converts the divergent beam of PC (b) into a parallel one, without changing the convergence of the parallel beam of PC (a) (the +1 and 0 orders of diffraction take place, respectively). The interferometer operates in the amplitude-division mode, with equal path lengths for both PCs. This feature essentially reduces the requirements on the coherency of the initial radiation.

The presented interferometer can be considered as an analog of the above mentioned triple Laue-case (LLL) interferometer. In the case of the LLL-interferometer, the actions of the splitter and the analyzer are based on different orders of the Bragg diffraction, in our case – on the different orders of diffraction on the FZP. In the case of the LLL-interferometer the blocks of the interferometer change the direction of propagation of the parallel beams, in our case they change the convergence of beams (a parallel beam is converted into converging and diverging in parallel). Both interferometers operate in the mode of amplitude-division, with equal path lengths for both PCs.

### 3. Defocused interferometers

The presented interferometer allows both longitudinal and transverse variety of defocusing.

#### *3.1. The longitudinal defocusing*

If the third block of the interferometer (analyzer) is shifted along the optical axis to the left at a distance $\Delta f \ll F$, then the PC (b) is converted from the plane wave into the slightly diverging wave with an imaginary source located to the left of the analyzer at a distance $L_3 = F(F - \Delta f)/\Delta f$, as shown in Fig. 2a. As a result of the interference of this beam with a parallel beam of PC (a), an interference pattern is formed in the form of concentric semicircles with radii

$$\rho_n = \sqrt{2\lambda p n} \quad (n = 1,\ 2, \dots). \tag{2}$$

Here $p = L_3 + f + \Delta f = F^2/\Delta f - F + f + \Delta f$ is the distance from the mentioned imaginary source to the image plane.

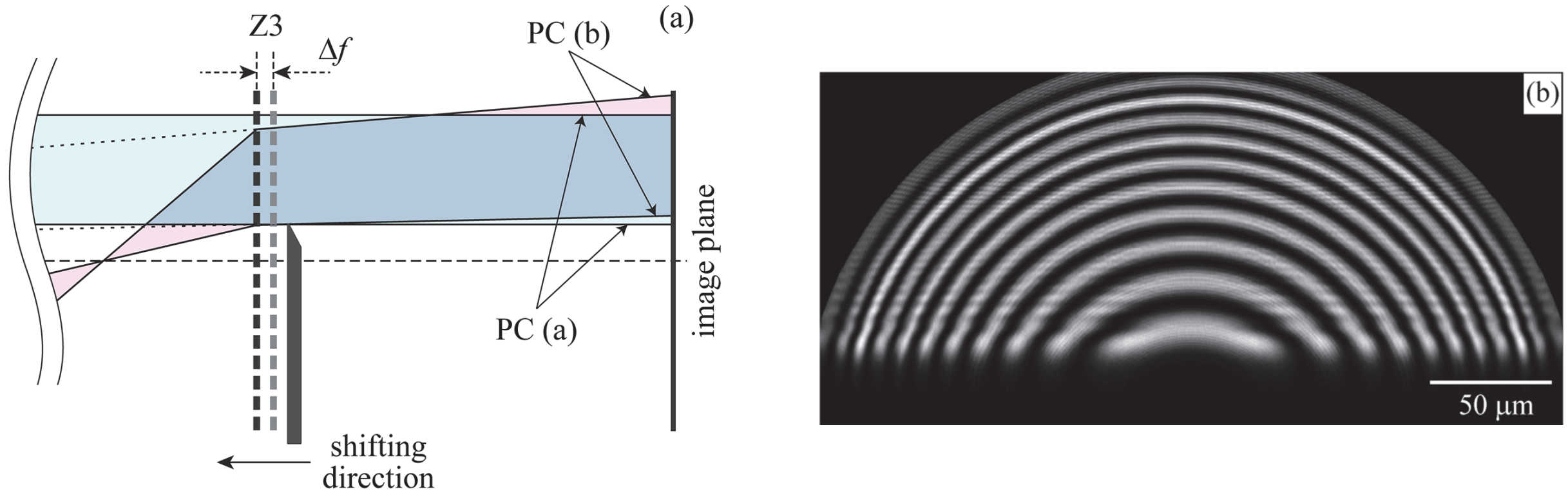

**Fig. 2.** (a) ray path and (b) numerically simulated interference pattern of the longitudinal defocused interferometer when the analyzer is shifted to the left. The dashed lines show the continuations of the rays of PC (b), which are intersecting at the imaginary source.

If the analyzer is shifted to the right, PC (b) is converted into a slightly converging beam, which leads to a similar interference pattern, with the parameter $p = |F^2/\Delta f + F - f + \Delta f|$.

The numerically simulated interference pattern for a left-shifted analyzer is shown in Fig. 2b. In addition to the clearly visible semirings, distortions in the form of horizontal lines are also noticeable in the figure, which are probably the result of X-ray diffraction on the knife-edges.

The carried out numerical simulations are based on the integration of the Helmholtz equation by the 2D Fourier-transform method [12]. Objects, intersecting the PC (in this case the FZP and knife-edges but afterward also the phase shifter and the testing object) are considered to be flat, and described by a complex amplitude transmission coefficient, which is a function of two coordinates perpendicular to the optical axis. Such a simplified approach is often used in hard X-ray optics. The reason is the small difference in the refractive index of substances from the unity and the small divergence of X-rays in the interferometer. The calculations are performed for $CuK_{\alpha1}$ radiation ($\lambda$ = 1.54 Å). The FZPs focal length is $F$ = 20 cm, and the number of Fresnel zones – $N$ = 760. Under these conditions, the FZP radius is $R$ = 153 μm and the outermost zone width – $\Delta r_N = 0.1$ μm. The etching depth of the zone structure of silicon FZPs was chosen 4.07 μm, which leads to a phase jump at the zone boundaries of 0.4π rad, and, accordingly, to the calculated value of the first-order focusing efficiency $\eta_1 \approx 14\%$ (without taking into account the absorption of radiation on the FZP substrate). The distance of the knife-edges from the optical axis is $d$ = 0.2$R$ = 30.6 μm, the interferometer–image plane distance for non-defocused interferometer – $f = 4F = 80$ cm, and the shifting of the analyzer is $\Delta f$ = 6.0 mm.

### *3.2. The transverse defocusing*

If the analyzer is shifted perpendicular to the optical axis, then the PC (b) will tilt towards the shifting direction by an angle $\varphi = \tau/F$, where $\tau$ is the analyzer shifting distance. As a result, an interference pattern is formed on the detector, in the form of parallel lines oriented perpendicular to the direction of the analyzer shift and with a period $l_{\mathrm{per}} = \lambda/\varphi = \lambda F/\tau = \Lambda^2/\tau$. Here $\Lambda = \sqrt{\lambda F}$ is the radius of the first zone of the FZP. The ray paths and numerically simulated interference pattern for this case are shown in Fig. 3.

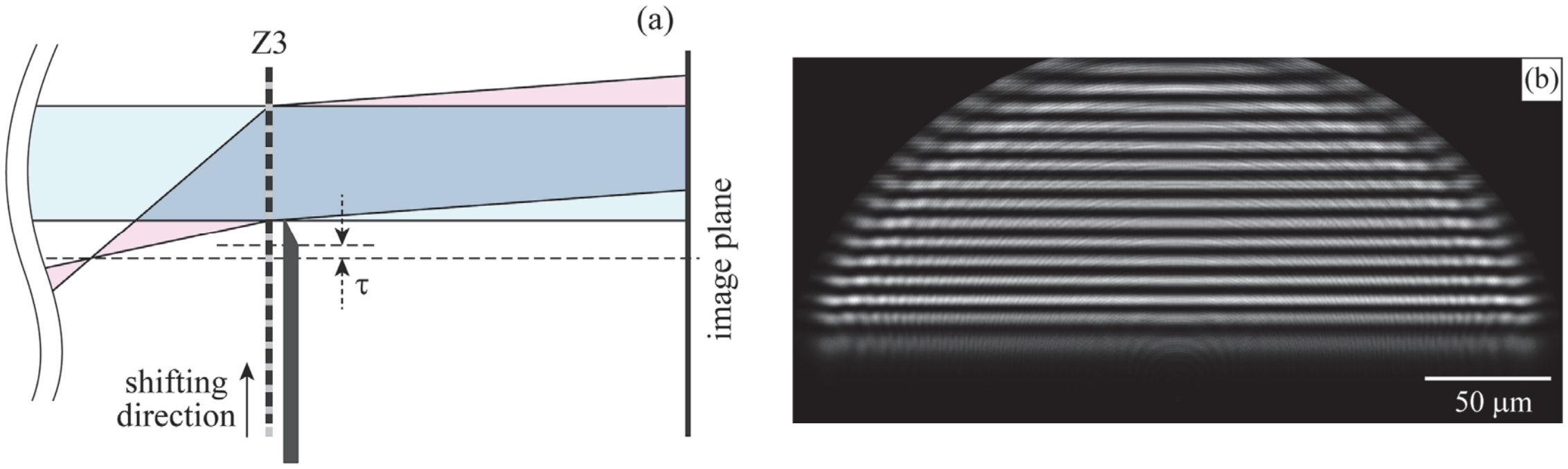

**Fig. 3.** (a) ray path and (b) numerically simulated interference pattern of the transverse defocused interferometer. The analyzer shifting distance: $\tau$ = 4 μm. The values of the other parameters are the same as in the case of Fig. 2b.

## 4. Phase-Contrast Imaging in the case of Initial Plane Wave

For phase-contrast imaging, the test phase object (an object of light elements, such as biological soft tissues), is located in the second inter-block space of the interferometer, as shown in Fig. 4. As can be seen from the figure, the second FZP, in addition to the mirror, plays the role of a condenser, concentrating the object wave on the test object. This is important both for the gain in intensity and for balancing the intensities of interfering waves (the object and the reference waves). The third FZP, in addition to the analyzer, plays the role of an objective, forming a sharp image of the test object on the image plane. The distances of the object plane from the focus point of the object wave ($q$) and the analyzer from the image plane ($f$) are connected by the law of image formation on a thin lens:

$$\frac{1}{F} = \frac{1}{F+q} + \frac{1}{f}, \tag{3}$$

which gives

$$q = F/G, \quad f = F(G+1),$$

where $G = f/(F+q)$ is the image magnification coefficient. In order to increase the angular aperture of the objective lens, and, hence, the device resolution, the distance of the second knife-edge from the optical axis ($d_2$) is reduced compared to the same distance for the first knife-edge ($d_1$). The phase-shifter is located across the reference wave, behind the second FZP.

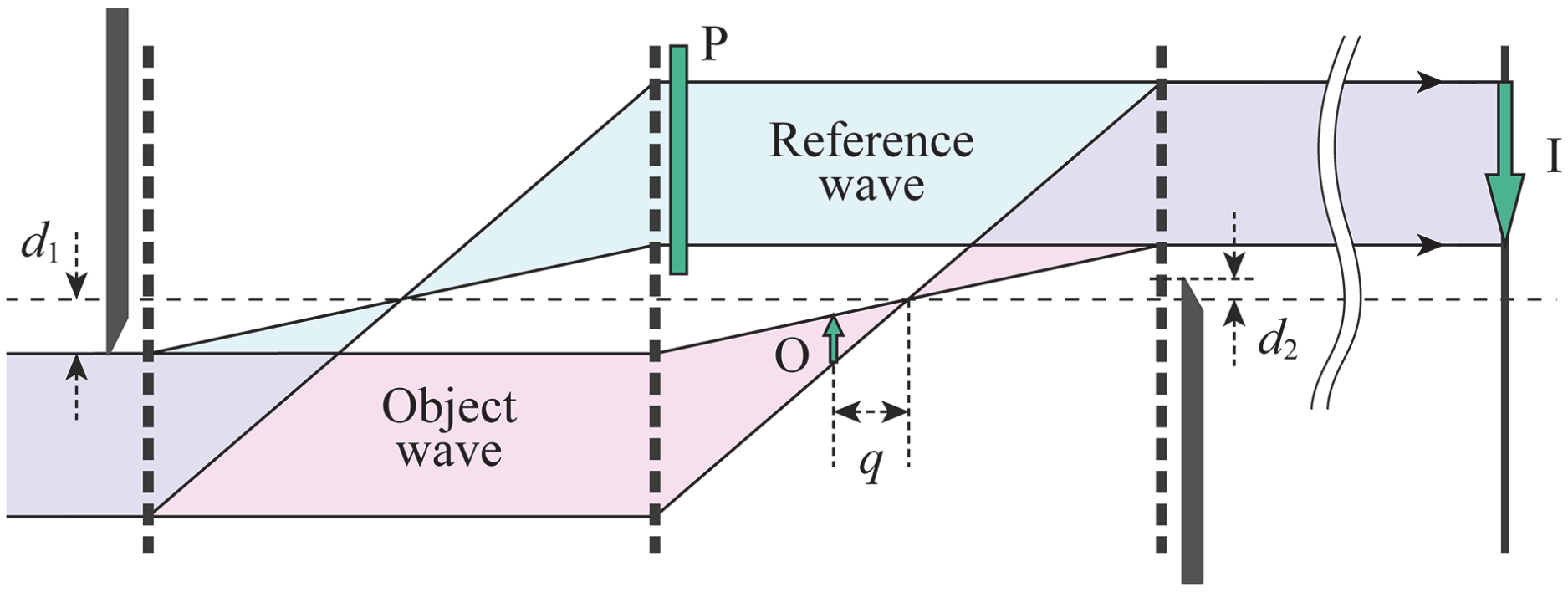

**Fig. 4.** Experimental setup of the phase-contrast image in the case of an initial plane wave. O is the test object, I – the image of the test object, and P – the phase-shifter.

As a result of the interference of the object and reference waves, the intensity distribution in the image plane is represented by the expression

$$I(\mathbf{r}) = a + b\cos^2(\Delta\varphi(\mathbf{r}')/2). \tag{4}$$

Here

$$a = \left(\sqrt{I_{\text{obj}}} - \sqrt{I_{\text{ref}}}\right)^2, b = 4\sqrt{I_{\text{obj}} I_{\text{ref}}}, \Delta\varphi(\mathbf{r}') = \varphi(\mathbf{r}') - \varphi_{\text{ps}},$$

$I_{\text{obj}}$ and $I_{\text{ref}}$ are the intensities of the object and reference waves, respectively, $\varphi(\mathbf{r}')$ – the phase-shift of the object wave introduced by the test object at the point $\mathbf{r}'$ on the object-plane, $\varphi_{\text{ps}}$ – the phase-shift of the reference wave introduced by the phase-shifter, $\mathbf{r}'$ and $\mathbf{r}$ are radius vectors in the image and object planes, respectively (the origins of these vectors are at the intersections of the optical axis with the mentioned planes). In (4), the vectors $\mathbf{r}'$ and $\mathbf{r}$ are connected by the condition $\mathbf{r} = -G\mathbf{r}'$.

In the case of $\varphi_{\text{ps}} = -\pi$, the expression (4) is converted to the form

$$I(\mathbf{r}) = a + b\sin^2(\varphi(\mathbf{r}')/2),$$

which represents a one-to-one mapping between $\varphi$ and $I$, for the tested-objects that are small enough that $-\pi < \varphi(\mathbf{r}') \leq 0$.

The numerical simulation of the phase-contrast image is carried out according to the presented scheme. A silicon wafer with etched grooves forming a rectangular grid is considered as a test object (see Fig. 5). The grid step is 956 nm, the width and depth of the grooves are 383 nm and 3.06 μm, respectively. The phase shift of the phase-shifter is chosen as $\varphi_{\text{ps}} = \varphi_0 - \pi$, where $\varphi_0$ is the phase shift of the silicon wafer without etched grooves. Such a choice of the phase-shifter results in the imaging of grooves in the form of light stripes on a dark background. The numerically simulated image of the test-object is shown in Fig. 6. In addition to a clear image of the test object, distortions in the form of semicircles and horizontal stripes are also noticeable in the figure. The author considers these distortions as a result of x-ray diffraction on the edges of knives and FZPs. They are more expressed compared with analogous distortions in the cases of defocused interferometers (see Figs. 2b and 3b). This is due to the low depth of the grooves of the test object (in the considered case, the beams' phase shift caused by the grooves is only 0.3 rad). As a result, the beams in the image plane superposed with a large phase difference and mainly canceled each other, which increases the contrast of distortions against the background of the interference pattern.

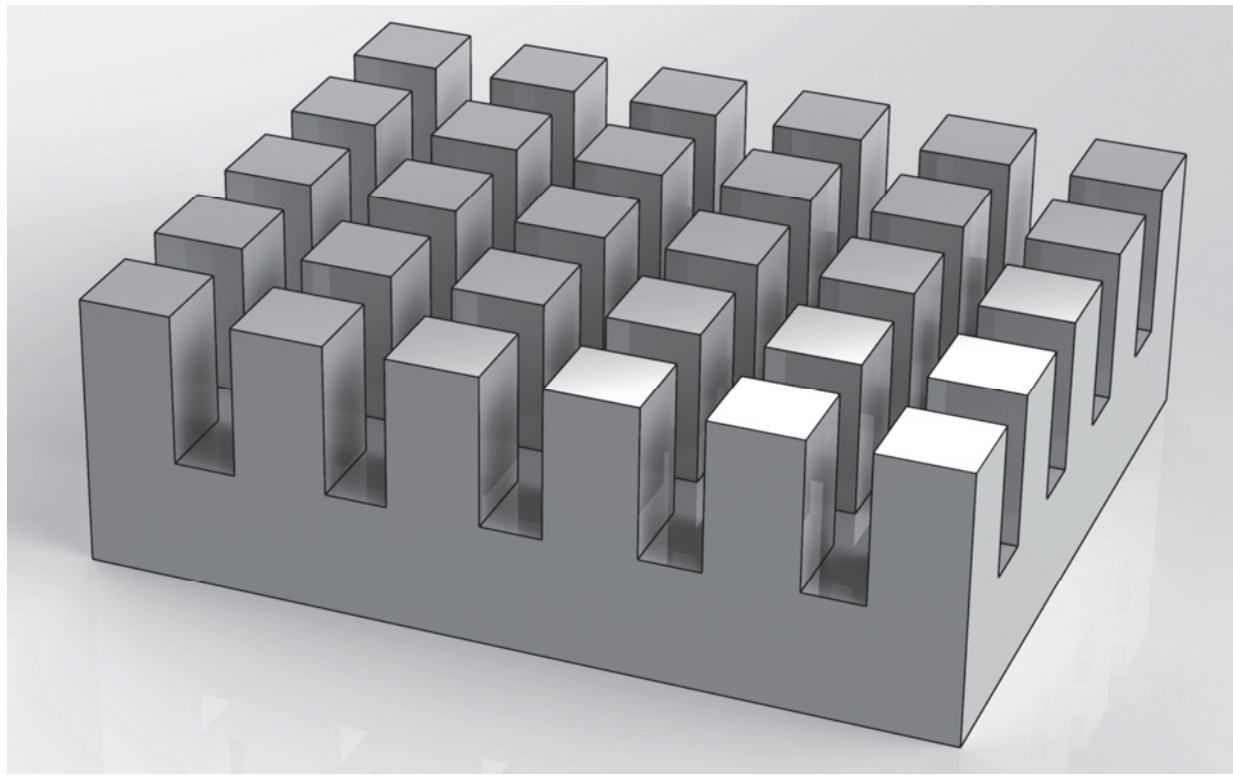

**Fig. 5.** The structure of the test object.

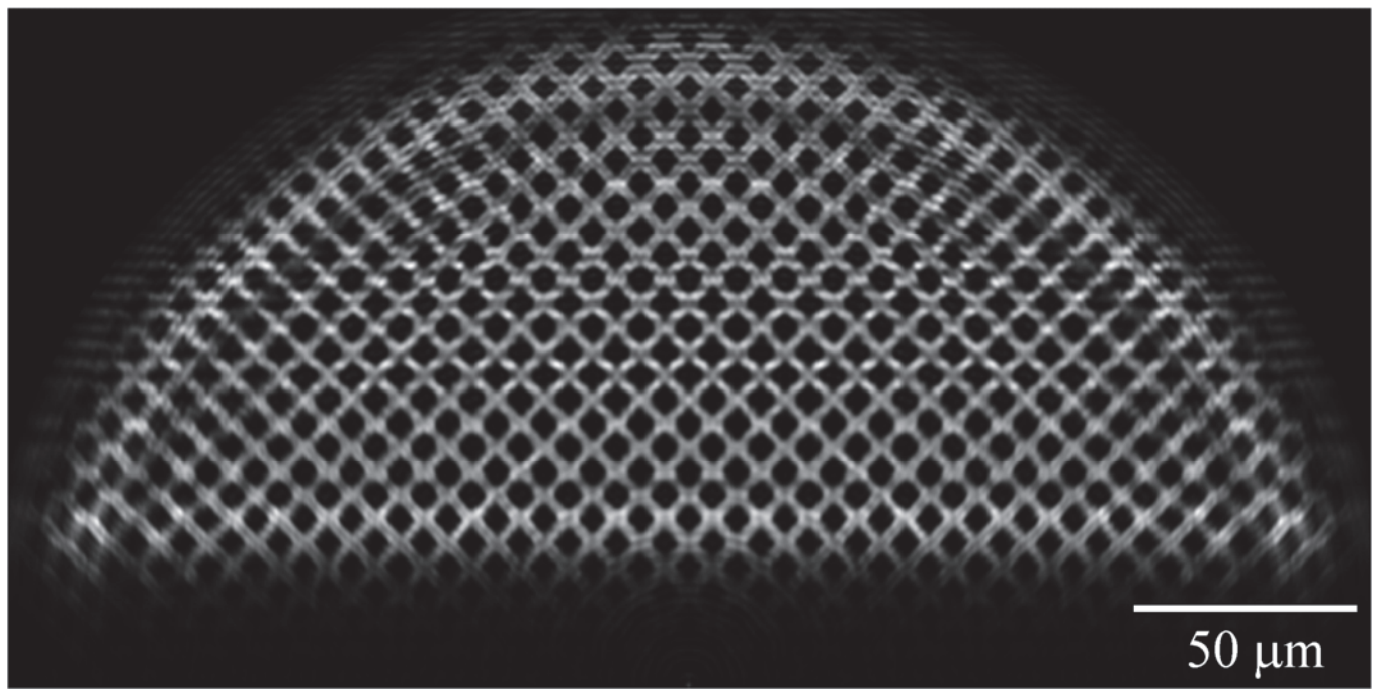

**Fig. 6.** Numerical simulation of the phase-contrast imaging in the case of an initial plane wave. The radiation wavelength and FZP parameters are the same as in the case of above considered defocused interferometers. The distances of the knife-edges from the optical axis are $d_1$ = 23.91 μm and $d_2$ = 4.78 μm for the first and second knife-edges, respectively. The image magnification factor is $G = 8$, which corresponds to the case of $q = F/G = 2.5$ cm and $f = (1 + G)F = 1.8$ m.

### 5. Phase contrast imaging using laboratory sources of X-ray radiation

So far, the case of the initial plane-wave X-ray radiation has been considered. This simplified approach is applicable for sources located far enough from the interferometer, which is generally achievable with synchrotron radiation sources. Let us now consider the case when the distance of a radiation point source from the interferometer is small enough to take into account the curvature of the wave-front of the radiation incident on the interferometer. The scheme of the interferometer modified for this case and the ray paths in it are shown in Fig. 7. The values of the main parameters of the scheme are selected according to the following conditions:

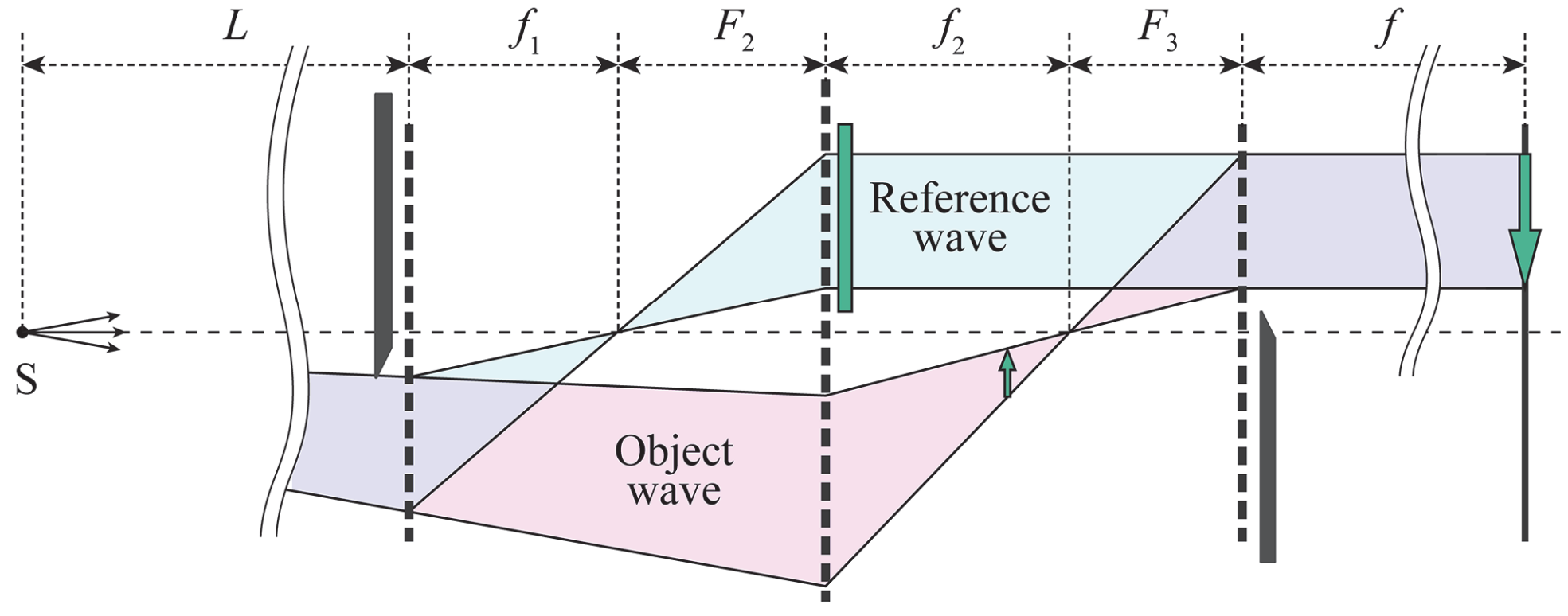

**Fig. 7.** Schematic diagram of the interferometer modified for the case of point source initial radiation. S is the X-ray radiation source and $L$ – the distance of the interferometer from the X-ray source.

$$F_2 = \frac{L}{L-F_1} F_1, \qquad F_3 = F_1,$$
$$R_2 = \frac{L+F_1}{L-F_1} R_1, \qquad R_3 = R_1, \tag{5}$$
$$f_1 = F_2, \qquad f_2 = \frac{L+F_1}{L-F_1} F_1.$$

Here $L$ is the source–interferometer distance, $F_i$ and $R_i$ are the focal length and radius of the $i$-th FZP, respectively. Due to the conditions (5), the interferometer still operates in the amplitude division mode, but now the path lengths of the interfering beams differ from each other. The maximum path difference is

$$\Delta l = \lambda N_1 \gamma, \tag{6}$$

where $N_1$ is the number of Fresnel zones of the first FZP, $\gamma = 2F_1/(L - F_1)$. The requirement for monochromaticity of the initial radiation corresponding to this path difference is:

$$\frac{\Delta\lambda}{\lambda} < \frac{\lambda}{2\Delta l} = \frac{1}{2N_1\gamma}. \tag{7}$$

For typical experimental conditions ($N{\sim}100 \div 1000$, $L{\sim}10F_1$), the requirement (7) becomes $\Delta\lambda/\lambda < 2.25 \times 10^{-3}$, which is quite achievable with modern laboratory sources of hard X-ray radiation.

The result of the numerical simulation of a phase contrast image using the presented above "modified" interferometer and a point-source of monochromatic X-ray radiation is shown in figure 8.

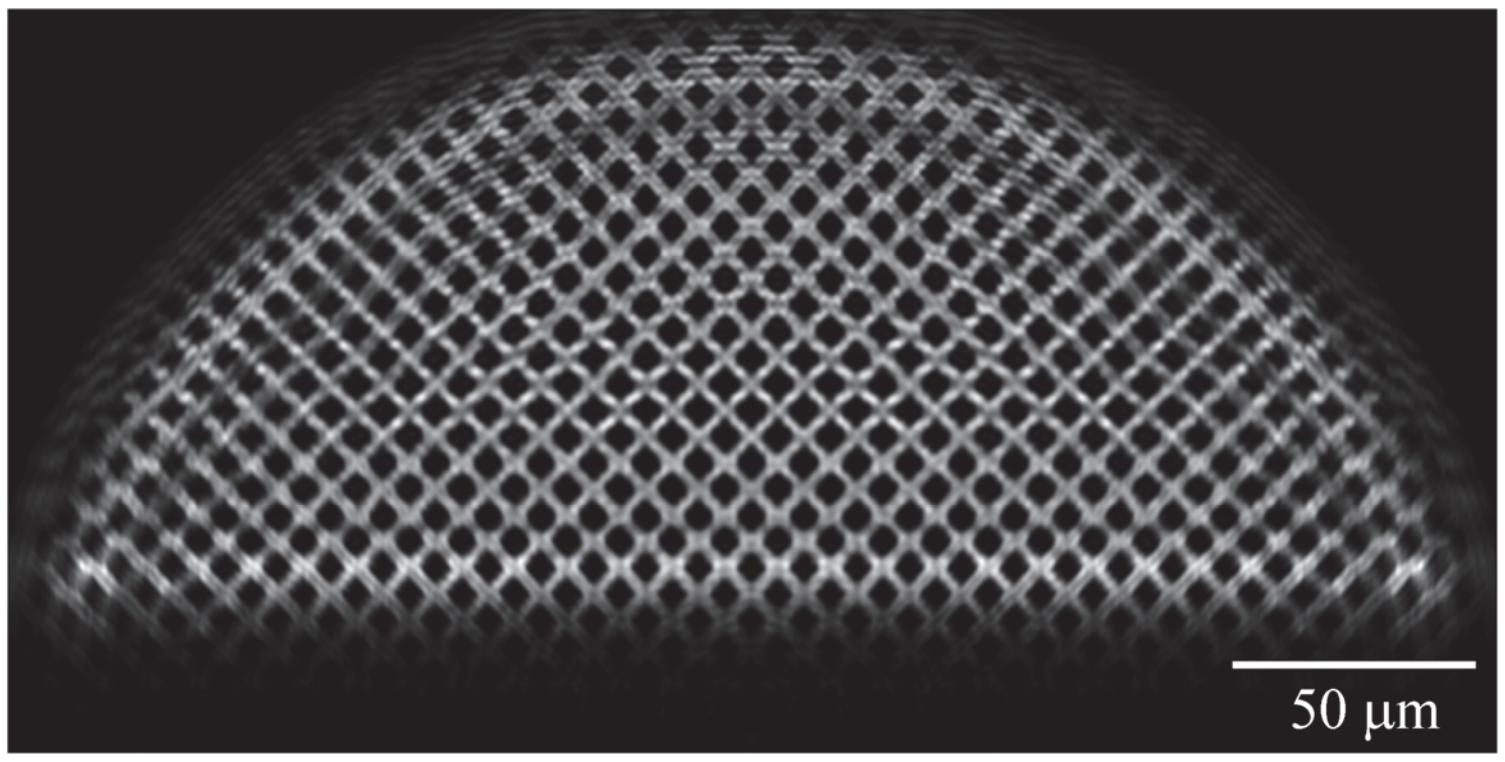

**Fig. 8.** Numerical simulation of the phase-contrast imaging using the modified interferometer in the case of a point-source monochromatic initial radiation. The radiation source is located on the optical axis at a distance $L = 10F_1 = 2$ m from the interferometer. The values of other parameters of the interferometer are the same as in the previous case, with the exception of the characteristics of the second FZP and interblock distances. The latter are determined according to (5) and are: the focal length and radius of the second FZP, respectively, are $F_2 = 22.2$ cm, and $R_2 = 187$ μm, the interblock distances are equal to each other and are: $f_1 + F_2 = f_2 + F_3 = 44.4$ cm.

In order to study the influence of limited coherence of the initial radiation on the image quality, calculations were also performed for spatially extended sources. The source is considered to consist of independently emitting points. The calculations were performed for both monochromatic and quasi-monochromatic radiation. In the second case, aiming to reduce the CPU time, the integration over the surface of the X-ray source was performed not over the uniformly distributed points, as in the first case, but only over the points on the diagonals of the square source. The radiation source size was chosen 190×190 μm$^2$, and the non-monochromaticity in the second case – $\Delta\lambda/\lambda = 2.2\times10^{-3}$.

As can be seen from the results of numerical simulations (see Figs. 9a, 9b), the limited spatial coherence of the initial radiation does not noticeably worsen the image quality. Moreover, it leads to the suppression of diffraction distortions, which are present in images with spatially coherent sources (Fig. 6 and 8). The

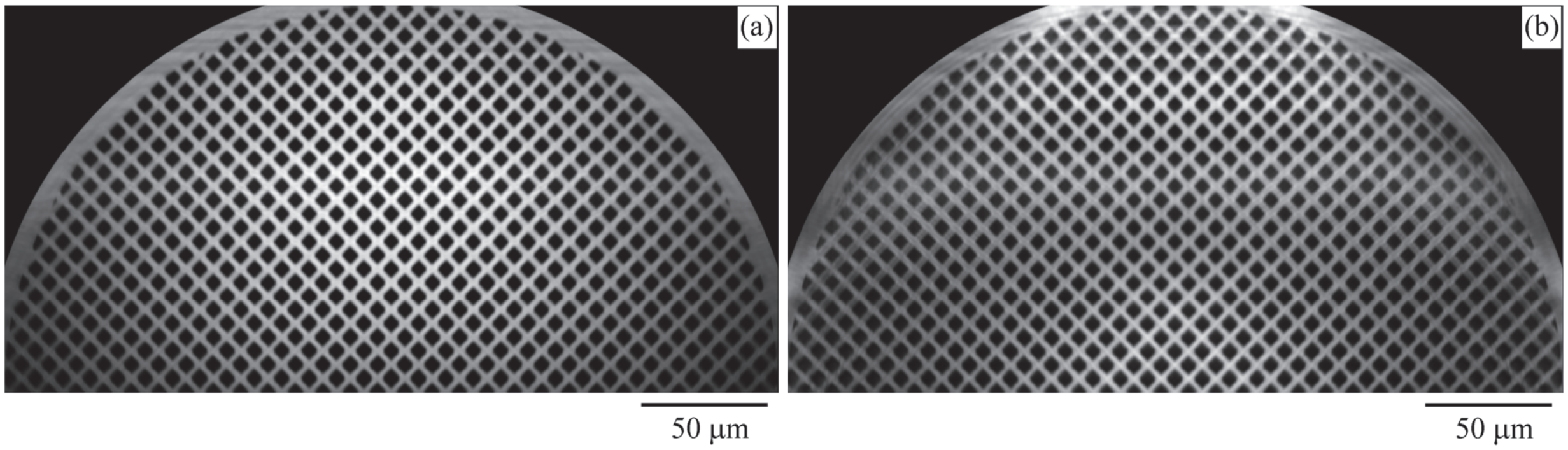

**Fig. 9.** Numerical simulation of the phase-contrast imaging using the modified interferometer in the case of an extended source of (a) monochromatic and (b) quasi-monochromatic radiation.

author believes that insignificant distortions in the case of a quasi-monochromatic source are a result of the above-mentioned decrease in the number of integration points

Note, that the mentioned characteristics of radiation sources are quite achievable for modern hard X-ray laboratory sources.

## 6. Conclusion

The possibility of hard X-ray phase-contrast imaging using a three-block FZP interferometer is considered by a numerical simulation. The mentioned interferometer operates in amplitude-division mode and does not impose strict requirements on the coherence of the initial radiation. The use of focusing elements as blocks of the interferometer makes it possible to obtain an image of an object with optical magnification. Note also, that a concentration of the light falling on the test-object takes place. This leads to intensity gain, which is especially important in the case of laboratory X-ray sources.

The phase-contrast imaging at a limited source-interferometer distance is considered both in the case of monochromatic point-source radiation and taking into account the size and non-monochromaticity of the source. The numerical simulation shows good quality of test-object imaging at a radiation source with sizes of $190\times190$ $\mu m^2$, non-monochromaticity $\Delta\lambda/\lambda = 2.2\times10^{-3}$ and source–interferometer distances of 2 m. Moreover, the limited spatial coherence of the initial radiation even improves the quality of imaging, by suppressing diffraction distortions.

Based on the above, it is assumed that the presented experimental setup can be used for phase-contrast imaging in the optical magnification mode using laboratory sources of hard X-ray radiation.

REFERENCES

1. U. Bonse and M. Hart, "An X-Ray Interferometer", *Appl. Phys. Lett.*, 1965, vol. 6, p. 155.
2. U. Bonse and M. Hart, "Principles and Design of Laue-Case X-Ray Interferometers", *Zeitschrift für Physik*, 1965, vol. 188, p. 154.
3. A. Momose, "Demonstration of phase-contrast X-ray computed tomography using an X-ray interferometer", *Nucl. Instrum. Methods Phys. Res., Sect. A*, 1995, vol. 352, p. 622.
4. M. Takeda, H. Ina, and S. Kobayashi, "Fourier-transform method of fringe-pattern analysis for computer-based topography and interferometry", *J. Opt. Soc. Am.*, 1982, vol. 72, p. 156.
5. J.H. Bruning, D.R. Herriott, J.E. Gallagher, D.P. Rosenfeld, A.D. White, and D.J. Brangaccio, "Digital wavefront measuring interferometer for testing optical surfaces and lenses", *Appl. Opt.*, 1974, vol. 13, p. 2693.
6. L.A. Haroutunyan, "X-ray Phase Contrast Imaging with Optical Magnification Using Three-Block Interferometer with Bi-Level Fresnel Zone Plates", *J. Contemp. Phys.*, 2016, vol. 51, p. 284.

7. D. Attwood, Soft X-Rays and Extreme Ultraviolet Radiation. Principles and Applications, Cambridge University Press, 1999.
8. T. Koyama, Y. Kagoshima, I. Wada, A. Saikubo, K. Shimose, K. Hayashi, Y. Tsusaka, and J. Matsui. “High-Spatial-Resolution Phase Measurement by Micro-Interferometry Using a Hard X-Ray Imaging Microscope”, *Jpn. J. Appl. Phys.*, 2004, vol. 43, p. L421.
9. T. Wilhein, B. Kaulich, and J. Susini, "Two zone plate interference contrast microscopy at 4 keV photon energy", *Optics Communications*, 2001, vol. 193, p. 19.
10. T. Koyama, T. Tsuji, K. Yoshida, H. Takano, Y. Tsusaka, and Y. Kagoshima, "Hard X-Ray Nano-Interferometer and its Application to High-Spatial-Resolution Phase Tomography", *Jpn. J. Appl. Phys.*, 2006, vol. 45, p. L1159.
11. L.A. Haroutunyan, "Amplitude-Division Type X-Ray Interferometer Based on Bi-Level Fresnel Zone Plates", *J. Contemp. Phys.*, 2015, vol. 50, p. 292-295.
12. J. Goodman, Introduction to Fourier Optics, New York: McGraw-Hill, 1996.